 \documentclass{sig-alternate-2013}

\usepackage{url}
\usepackage{hyperref}
\usepackage{tikz}
\usepackage{graphicx}
\usepackage{subcaption}
\usepackage{caption}
\usepackage{bbm}
\usepackage{algorithm}
\usepackage{algpseudocode}
\usepackage{mathtools}
\usepackage[utf8]{inputenc} 
\usepackage[T1]{fontenc}
\usepackage[english]{babel}
\usepackage[autostyle]{csquotes} 
\newcommand{\argmin}{\arg\!\min}
\newtheorem{mydef}{Definition}
\newtheorem{mylem}{Lemma}

\newcommand{\pushright}[1]{\ifmeasuring@#1\else\omit\hfill$\displaystyle#1$\fi\ignorespaces}
\renewcommand{\algorithmicrequire}{\textbf{Input:}}
\renewcommand{\algorithmicensure}{\textbf{Output:}}
\newcommand\myeq{\stackrel{\mathclap{\normalfont\mbox{def}}}{=}}

\begin{document}

%
\permission{Copyright is held by the International World Wide Web Conference Committee (IW3C2). IW3C2 reserves the right to provide a hyperlink to the author's site if the Material is used in electronic media.}
\conferenceinfo{WWW 2016,}{April 11--15, 2016, Montr\'eal, Qu\'ebec, Canada.} 
\copyrightetc{ACM \the\acmcopyr}
\crdata{978-1-4503-4143-1/16/04. \\
http://dx.doi.org/10.1145/2872427.2882972}

\clubpenalty=10000 
\widowpenalty = 10000

\title{In a World That Counts: Clustering and Detecting \\Fake Social Engagement at Scale}

%
%
%
%
%

\numberofauthors{4} 
%

\author{
%
%
\alignauthor
Yixuan Li\titlenote{Work done while interning at Google Inc.} \\
\affaddr{Cornell Unversity}\\
\affaddr{Ithaca, NY 14853, USA}\\
\email{yli@cs.cornell.edu}
\alignauthor
Oscar Martinez \\
\affaddr{Google Inc.}\\
\affaddr{Mountain View, CA}\\
 \email{omartinez@google.com}
\alignauthor Xing Chen\\
\affaddr{Google Inc.}\\
\affaddr{Mountain View, CA}\\
\email{chenxing@google.com}\\
\and  
\alignauthor Yi Li\\
\affaddr{Google Inc.}\\
\affaddr{Mountain View, CA}\\
\email{yiyili@google.com}       
\alignauthor
John E. Hopcroft \\
\affaddr{Cornell University}\\
\affaddr{Ithaca, NY 14853, USA}\\
\email{jeh@cs.cornell.edu}\\
}



\maketitle
\begin{abstract}
How can web services that depend on user generated content discern fake social engagement activities by spammers from legitimate ones? In this paper, we focus on the social site of YouTube and the problem of identifying bad actors posting inorganic contents and inflating the count of social engagement metrics. We propose an effective method, \textsc{Leas} ({\em Local Expansion at Scale}), and show how the fake engagement activities on YouTube can be tracked over time by analyzing the temporal graph based on the engagement behavior pattern between users and YouTube videos. With the domain knowledge of spammer seeds, we formulate and tackle the problem in a semi-supervised manner --- with the objective of searching for individuals that have similar pattern of behavior as the known seeds --- based on a graph diffusion process via local spectral subspace. We offer a fast, scalable MapReduce deployment adapted from the localized spectral clustering algorithm. We demonstrate the effectiveness of our deployment at Google by achieving a manual review accuracy of 98\% on YouTube Comments graph in practice. Comparing with the state-of-the-art algorithm \textsc{Copy}\textsc{Catch}, \textsc{Leas} achieves 10 times faster running time on average. \textsc{Leas} is now actively in use at Google, searching for daily deceptive practices on YouTube's engagement graph spanning over a billion users.  

\end{abstract}



\keywords{Fake social engagement; Anomaly detection; 
Local spectral clustering; Seed expansion; Social networks }

\section{Introduction}
Every day people generate a large amount of comments on YouTube but not all of those engagement activities are real. Bad actors have been trying to game the system by posting inorganic contents and inflating the count of social engagement metrics.

We consider any practice that attempts to post fake contents, or artificially inflate the number of YouTube engagement metrics through the use of automated means or as a marketplace, as illegitimate activity. Generally speaking, any engagement activity in online social media that does not reflect user's genuine interest can be viewed as {\em fake social engagement}. 

The issue of fake social engagement came into being partly due to that third-party businesses attempt to boost YouTube video engagement metrics in order for promoting contents and increasing popularity. At Google, we have seen attackers attempting to take advantage of the YouTube community by using a variety of deceptive practices \cite{youtube:policy:spam}, including malware, fake accounts, artificial traffic spam and comment spam. Among the various forms of spam activity, fake social engagement has become the most frequently seen yet hardest to detect practice. In particular, we have discovered that abusive YouTube comments  have evolved from traditionally {\em explicit} spammy-like (e.g. linked with bad URLs or associated with obvious advertisement), to a more {\em insinuated} outlook that makes them difficult to discern from those organic comments. For instance, one common type of fake YouTube comments comprises text pieces such as ``{\em cool }", ``{\em oh}", which has made approaches largely basing on text features and bad URL detection insufficient in such scenario. 

At Google, we note the importance of keeping the service free of fake engagement activities that may potentially spoil the online social ecosystem. As the YouTube official policy guide on subscription \cite{youtube:policy} states, for example: 

\blockquote{\em 
 Subscribing to a channel creates a relationship between a content creator and a content consumer; the creator keeps making great videos, and the consumer keeps watching, like-ing, and commenting. We take this relationship seriously. A subscription is a user-initiated pledge of support to a YouTube channel; this means that a real human being wants this channel's content in their feed every day. The amount of users subscribed to a YouTube channel should be a metric that reflects genuine interest in that channel, not a gauge of automated or falsified activity.}
  And we believe that such policy does not only apply to YouTube Subscribes, but can also  be extended to other engagement activities such as Comments as well. As a matter of fact, YouTube is far from the only social media facing the challenge of keeping its service free of deceptive practices. Twitter Followers and Facebook Likes are all buyable by the thousand online \cite{de2014paying}, for example. 



To address these issues, we study the temporal engagement activity patterns on YouTube, making use of anonymized aggregate daily logs of YouTube Comments. We create the engagement relationship graph by taking account the frequency of common engagement activities shared between two individuals within a short period of time. The engagement graph allows us to detect {\em orchestrated actions} by sets of users which have a very low likelihood of happening spontaneously or organically. Such behavior of groups of users acting together on the same videos or channels at around the same time, is also known as {\em lockstep behavior} \cite{beutel2013copycatch}. 

To detect the lockstep behavior on YouTube, we take a semi-supervised learning approach, making use of existing known abusive accounts as {\em seeds}. We demonstrate an effective method, {\em \textsc{Leas} (Local Expansion at Scale)}\footnote{Interestingly, the word ``leas" has the meaning of ``well-being" in old Irish.}, in detecting deceitful user engagement based on the local spectral graph diffusion \cite{li2015uncovering}. Local spectral method has substantial advantage over traditional spectral techniques because of its capability in prioritizing and finding clusters {\em only} near a local region of the engagement graph surrounding the seed. Specifically, \textsc{Leas} searches for clusters consisting of suspicious nodes with similar pattern of behavior as the given seeds.  We show \textsc{Leas} is scalable to massive datasets, with a straightforward adaption to the MapReduce implementation. Moreover, the MapReduce deployment has the same performance guarantee as the serialization since each diffusion procedure is performed locally. By clustering YouTube users based on their engagement behavior pattern, \textsc{Leas} can greatly expand the coverage of daily fake engagement take-down volume on YouTube. Our approach can be extended to many other settings including Twitter followers, Amazon product reviews and Facebook Likes etc. 

The ultimate goal of our research effort is to help improve social media environment as well as  user experience, and to ensure an online world where contents and clicks can be translated into genuine and meaningful interactions. Toward achieving the goal, this paper offers a number of contributions listed in the following:

\begin{enumerate}
\item {\bf Problem Formulation:} We provide a novel problem formulation --- a semi-supervised learning problem based on the local spectral graph diffusion --- to a real-world challenge realized at Google and relevant in many online settings. One advantage of our setting is its full generality. That is, it is applicable for any similarity-based graph without much need for customization.
\item {\bf Algorithm:} We offer a fast, scalable MapReduce implementation adapted from the localized spectral clustering algorithm \cite{li2015uncovering}. This is the first large-scale deployment of local spectral clustering to the best of our knowledge.
\item {\bf Behavioral Analysis:} We show comprehensive performance evaluations on \textsc{Leas} --- focusing on multitudes of different characteristics exhibited by abusive accounts compared to that of general population --- using both structural and contextual information. 
\end{enumerate}






The remainder of the paper is organized as follows. Section \ref{sec:related} describes related work on using graph-based approaches in detecting anomalies. In Section \ref{sec:formulation} and Section \ref{sec:methodology} we mathematically formulate the problem and describe how it can be solved in a semi-supervised learning framework. We introduce the YouTube engagement graph dataset in Section \ref{sec:graph}. A MapReduce implementation is discussed in Section \ref{sec:mapreduce}. Finally in Section \ref{sec:experiments} we offer experimental analysis, demonstrating the usefulness of our deployment at Google; and conclude our work in Section \ref{sec:conclusion}.

\section{Related Work} \label{sec:related}

Online spam activities are evolving as fast as the web services themselves. Abusive actions have been observed in a wide range of domains, including Email \cite{chirita2005mailrank}, web search \cite{becchetti2008link,castillo2007know,ntoulas2006detecting,wang2007spam,wu2006topical} and blogs \cite{kolari2006detecting}. In recent years, spam campaigns have also been prevalently emerging on major social media sites \cite{gao2010detecting,stringhini2010detecting}, with a diverse set of application targets spanning YouTube \cite{benevenuto2009detecting,o2012network}, Facebook \cite{beutel2013copycatch,cao2014uncovering,de2014paying}, Amazon \cite{lim2010detecting,mukherjee2012spotting}, Twitter \cite{benevenuto2010detecting,wang2010don}, eBay \cite{pandit2007netprobe}, and many others. 

A number of content-based spam detection strategies have been exploited in the past decade \cite{castillo2007know,ntoulas2006detecting,wang2007spam,wu2006topical}. Most of the proposed methods rely on extracting evidences from textual descriptions of the content, treating the text corpus as a set of objects with associated attributes, and applying classification method such as Support Vector Machine (SVM) \cite{joachims1998text} to detect spam \cite{heymann2007fighting,ntoulas2006detecting}. A few other more sophisticated methods also take into account the multimedia information such as image features \cite{mehta2008detecting,wu2005using}.

Content and link based approaches, however, can be infeasible in identifying fake social engagement when contextual information is unavailable, or faking to be organic-like. Many papers tend to devise feature-based classifiers incorporating various account-level as well as social relationship features \cite{benevenuto2009detecting,stringhini2010detecting,yang2014uncovering}. While these supervised training models are useful at depicting the spam strategy behind the observed temporal dynamics, it is nonetheless unclear how generalizable they are beyond the particular product or signal studied.  Furthermore, it is practically difficult to obtain large volumes of training data because manual labeling can be expensive. 
To this end, recent proposals based on {\em behavioral clustering} have demonstrated to be effective in spotting groups of users with similar behavior patterns in terms of engagement activities \cite{beutel2013copycatch,cao2014uncovering,jiang2014inferring,mukherjee2012spotting,wang2013you, stringhini2015evilcohort}. These methods often start with constructing a bipartite graph representing user-product engagement relationships. Various unsupervised clustering techniques (e.g., co-clustering \cite{beutel2013copycatch} and community detection \cite{stringhini2015evilcohort}) have been applied for detecting groups of actors with similar behavior. Below we highlight a few and illustrate how our work contributes to this line.

More related to our own work, Beutel  et al. \cite{beutel2013copycatch} investigated the problem of fake Page Likes on Facebook and observed a lockstep behavior pattern exhibited by spammers, where groups of users often acting together and Like-ing the same Pages in a loosely synchronized manner. Such collaborative spamming behavior was also observed in Twitter \cite{lee2012detecting} and also Amazon product reviews \cite{mukherjee2012spotting}, where paid groups of frequent fake review writers have been trying to promote or demote certain products on Amazon. The incremental work \cite{cao2014uncovering} further extended the \textsc{Copy}\textsc{Catch} approach \cite{beutel2013copycatch} to several other applications such as Facebook app install and Instagram follow. Note that our setting advances \cite{beutel2013copycatch} by making use of possible domain information in a semi-supervised manner; and our problem formulation is also complementary to \cite{mukherjee2012spotting} which required a large number of both positive and negative examples in a fully supervised manner. 

Our work builds on these papers, providing advances in two aspects: algorithmically, our clustering algorithm is operated in a fully localized fashion, which is efficient to compute and easily parallelizable; practically, our framework is fully generalizable, which can be extended to other behavioral clustering problems and applications without much need for customization. 


\section{Problem Formulation} \label{sec:formulation}

We now describe the mathematical formulation of our problem. We take a semi-supervised approach and define suspicious behavior in terms of graph structure and edge creation times. 

To make our problem definition more generalizable, we adopt the notions of {\em actor} and {\em target} in representing the entities involved in an engagement activity. For example, in the context of YouTube Comments, a target can be translated into a video.

In the following, we introduce two types of graph that can be created using the engagement activity information. A straightforward way is to build an {\em engagement bipartite graph} between the set of actors and the set of target, where we use edge to indicate the engagement timestamp. Since we are interested in clustering entities of actors, a more refined way would be to construct an {\em engagement relationship graph},
in which nodes consist of all the actors and two nodes share an edge if they have acted upon the same target(s). 


Mathematically, assuming we are provided with a set of actors, $\mathcal{V}=\{v_i\}_{i=1}^{|\mathcal{V}|}$ and a set of targets $\mathcal{Q}=\{q_j\}_{j=1}^{|\mathcal{K}|}$. We are also given a set of seeds $\mathcal{S}=\{s_r\}_{r=1}^{|\mathcal{S}|}$. Each engagement activity can be described by a tuple of $(v_i,q_j,t_{v_i\rightarrow q_j})$, where $t_{v_i \rightarrow q_j}$ records the timestamp at which actor $v_i$ acted on target $q_j$.    

\begin{itemize}
\item {\bf Engagement Bipartite:} We define $B = (\mathcal{V}, \mathcal{Q}, \mathcal{T})$ as a temporal engagement bipartite graph, where each timestamped edge $(v_i,q_j) \in \mathcal{T}$ records the time at which $v_i \in \mathcal{V}$ acted on $q_j\in \mathcal{Q}$. We further enforce the temporal constraint that all the actors acted on the targets in a 2$\Delta t$ time window, i.e., 
\begin{equation}
\exists t_r \in \mathbb{R}~\text{s.t.}~|t_r - t_{v_i\rightarrow q_j}| \le \Delta t~\forall v_i \in \mathcal{V}, q_j \in \mathcal{Q}
\end{equation}

\item {\bf Engagement Relationship Graph:} We define $G=(\mathcal{V},\mathcal{E},\mathcal{W})$ as a temporal engagement relationship graph. $\mathcal{E}$ is the edge set, where $(v_i,v_j) \in \mathcal{E}$ if actors $v_i$ and $v_j$ have acted upon the same non-empty set of target $\mathcal{Q}_{v_i,v_j} \subseteq \mathcal{Q}$, with weight denoted by $w_{v_i,v_j} \in \mathcal{W}$. Further details regarding the edge weight will be discussed in Section \ref{sec:graph}.
\end{itemize}

Throughout the paper, our methods and analysis will be focusing on the engagement relationship graph. And we will henceforth use the term {\em engagement graph} for brevity.  

\vspace{0.2em}
\noindent {\bf Given:} An engagement graph $G=(\mathcal{V},\mathcal{E},\mathcal{W})$ that models the intensity engagement relationship between nodes; and the seed set $\mathcal{S}$. 

\vspace{0.2em}
\noindent {\bf Output:} Accomplice clusters $\mathcal{C}_1$, $\mathcal{C}_1$,..., $\mathcal{C}_{|\mathcal{S}|}$ corresponding to each seed in the set $\mathcal{S}$. Each cluster consists of suspicious nodes with similar pattern of behavior as the given seed, which satisfy the definition of $[n,m,\rho,\Delta t]${\em -temporally approximate bipartite core} (T-ABC) given below.  
\begin{mydef}
We define an $[n,m,\rho,\Delta t]$-temporally approximate bipartite core (T-ABC) with respect to a given seed $s  \in \mathcal{S}$, as a set of actors $\mathcal{C'}\subseteq \mathcal{V}$ associated with a set of edges $\mathcal{E'} \subseteq \mathcal{E}$ such that
\begin{align} 
s &\in \mathcal{C'}\\
|\mathcal{C'}| &\ge n \\ 
|\mathcal{E'}| &\ge \rho \cdot \frac{n(n-1)}{2} \\
w_{v_i,v_j} &\ge m~\forall (v_i,v_j) \in \mathcal{E'}
\end{align}

\end{mydef}
Here we introduce the term $\rho\in [0,1]$ to relax the constraint in the original definition of $[n,m,\Delta t]$-temporally coherent bipartite core (TBC) in \cite{beutel2013copycatch}. We make such change since we find many loosely connected abusive clusters existing in practice. Relaxing the constraint enables us finding both tightly and loosely connected groups of suspicious actors.


\section{Semi-supervised learning via local spectral diffusion} \label{sec:methodology}


We use the semi-supervised learning method to tackle the problem of detecting the suspicious actor groups defined in previous Section. Graph-based learning approach can be viewed as a probability diffusion that propagates large values from a small set of nodes with known labels --- which are usually referred to as {\em seeds} in literature --- to the remaining nodes of the graph \cite{gleich2015using}. This type of approach typically starts with a graph and the labeled sample matrix $\mathbf{S}\in \mathbb{R}^{N \times K}$, where $N$ is the number of nodes in the graph and $K$ is the number of classes. $S_{i,j}=1$ if node $i$ is labeled with class $j$, and $S_{i,j}=0$ otherwise. 

A graph-based learning framework usually incorporates two essential parts. The first is  to produce an $N\times K$ matrix $\mathbf{Y}$ which {\em encodes} the probability for each unlabeled node to be in certain classes. Specifically, $Y_{i,j}$ should be large if node $i$ should be labeled as class $j$. In our problem setting of binary classification, the diffusion matrix can be reduced to a vector $\mathbf{y} \in \mathbb{R}^N$, where larger value indicates a higher possibility being labeled the same as the seeds. And the second key component is to {\em decode} the diffusion values in $\mathbf{y}$ into a predicted label based on some graph metric optimization criterion. In the following, we will provide details on both components of our learning algorithm.

\vspace{1em}
{\bf \noindent Local spectra vs. global spectra}

\noindent Spectral methods is one of the most widely used techniques for exploratory data analysis, with applications ranging from data clustering, image segmentation to community detection etc. Spectral clustering makes use of the first few singular vectors of the Laplacian matrix associated with a graph, which are inherently global quantities and may not be sensitive to very local information \cite{mahoney2012local}. For example, in the case when provided with domain knowledge about a target region in the graph, one might be interested in finding clusters {\em only} near the specified local region in a semi-supervised manner,  which might not be otherwise well captured by a method using global eigenvectors. Therefore, in the semi-supervised setting, our pioneer work on local spectral clustering  \cite{li2015uncovering} have substantial advantage over traditional spectral techniques, with the capability of prioritizing and learning more about a local region of the graph surrounding the seeds.     

\subsection{Degree-thresholded Sampling}
We apply a degree-thresholded sampling procedure using breadth-first-search (BFS) to get a small subgraph $G_s$ covering a local neighborhood region surrounding the seed. Starting from the given seed $s$, we take the set of frontier nodes --- except for those nodes with degree larger than $d_\text{max}$ --- into the subgraph node set and repeat the process until the size of the subgraph reaches the specified upper limit $N$. We enforce the degree thresholding to prevent including extremely high-degree nodes, which are less likely to be spammers\footnote{We set $d_\text{max}$ to be 500 by default. This is because the degree of most known spammer nodes is smaller than 500, as shown in Figure \ref{fig:degree}.}. In practice, we choose the parameter of $N$ to be at least several times larger than the maximum size of the cluster of interest $|\mathcal{C}_s|$, in order to capture as many nodes in the target group as possible.    


\subsection{Local Spectral Subspace}

Consider the subgraph graph $G_s$ extracted from the neighborhood surrounding the seed $s$. We define  the normalized adjacency matrix $\mathbf{ \bar A}_s$ of the graph $G_s$ as 
\begin{equation}\label{eqn:normalized_matrix}
\mathbf{ \bar A}_s \myeq \mathbf{D}_s^{-1/2}(\mathbf{A}_s+\mathbf{I}) \mathbf{D}_s^{-1/2},
\end{equation}
where $\mathbf{A}_s$ and $\mathbf{D}_s$ denotes the adjacency matrix and the diagonal degree matrix of $G_s$, respectively. Let $\mathbf{p_0}$ denote the initial probability vector with element 1 in the entry of the seed node and 0 elsewhere. We describe how to efficiently construct the local spectra by iteratively transforming the orthonormal basis starting with a {\em Krylov subspace} defined below. 
\begin{mydef}
The order-$l+1$ Krylov subspace generated by the matrix $\mathbf{A} $ and vector $\mathbf{p}_0$ is the linear spanned subspace defined by the probability vectors in $l$ successive random walks
\end{mydef}
\begin{equation}
\mathcal{K}_{l+1}(\mathbf{A},\mathbf{p}_0) = {\text{span}}~\Big(\mathbf{p}_0,\mathbf{A} \mathbf{p}_0,...,\mathbf{A}^l \mathbf{p}_0 \Big).
\end{equation}

\begin{figure}
\centering
\includegraphics[width=\columnwidth]{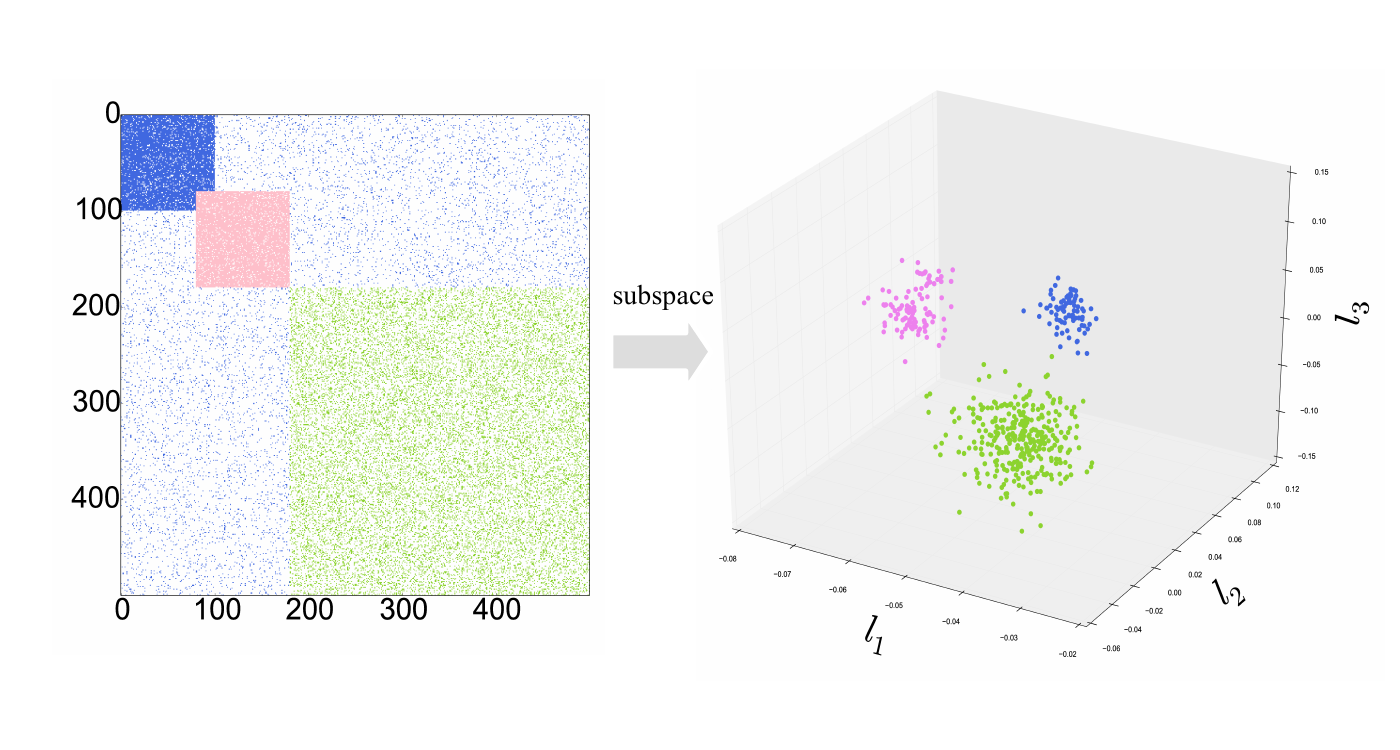}%
\caption{{\bf An example of local spectral subspace $\mathbf{V}_{3,3}$. The synthetic subgraph $G_s$ is generated with Erd\H{o}s-Rényi $G(n,p)$ model with background noise $p=0.05$. The spammer group $A$ and $B$ (denoted by blue and pink respectively) are of size 100 with edge probabolity $p=0.9$, with partially overlapped 20 nodes. The non-spammer group $C$ (denoted by the green color) has size 320 with $p=0.2$. The subspace is generated by Algorithm 1 starting from the seed with index 10 in the spammer group $A$. }}%
\label{fig:spectra}%
\end{figure}

In Algorithm 1, we briefly summarize the procedure of calculating the local spectral subspace from a specified seed. We start by calculating the initial invariant subspace $\mathbf{V}_{0,l}$, which is the orthonormal basis of $\mathcal{K}_{l+1}(\mathbf{A_\mathcal{S}},\mathbf{p}_0)$. And the local spectral subspace can be then obtained by iterating the process specified in \textsc{Line} 4-6 of Algorithm 1. The random walk step $k$ and subspace dimension $l$ are the key parameters in the local spectral clustering algorithm. Following the heuristics in 
\cite{li2015uncovering}, we set $k=3$ and $l=3$ respectively. Figure \ref{fig:spectra} \cite{li2015overlapping} shows an example local spectral subspace $\mathbf{V}_{3,3}$, generated from a synthetic graph with  Erd\H{o}s-R\'{e}nyi $G(n,p)$ model. The visualization demonstrates that local spectral subspace enables capturing the closeness of entities belonging to the same group.

\subsection{Learning via Local Spectral Diffusion} \label{section:mon}

With the local spectra $\mathbf{V}_{k,l}$, we then solve the following $\ell_1$ norm optimization problem.
\begin{align}
\min &~~~||\mathbf{y}||_1 \\
\text{s.t.} & ~~~\mathbf{y}=\mathbf{V}_{k,l}\mathbf{z}, \\
&~~~\mathbf{y} \ge \mathbf{0},\\
&~~~ \mathbf{y}(s) \ge 1,
\end{align}
where the objective function itself is a regularized term with sparsity penalty. Both $\mathbf{z}$ and $\mathbf{y}$ are unknown vectors.
The first constraint indicates that $\mathbf{y}$ is in the space of $\mathbf{V}_{k,l}$.
The element in $\mathbf{y}$ 
 indicate the likelihood for the corresponding node being labeled the same as seed $s$, which is non-negative.
The third constraint enforces that seeds are in the support of sparse vector $\mathbf{y}$. 

To interpret the optimization formulation from a geometric perspective, we are essentially seeking a sparse vector in the span of the local spectral subspace, such that the seed is in its support. In other words, the learning objective here is a locally-biased spectral program which enforces the solution to be well-connected with the known seed $s$. 

\floatname{algorithm}{Algorithm}
\renewcommand{\algorithmicrequire}{\textbf{Input:}}
\renewcommand{\algorithmicensure}{\textbf{Output:}}
\begin{algorithm}[htbp]
  \caption{\textsc{Local}\textsc{Spectral}($G_s,s$)}
  \begin{algorithmic}[1]
    \Require{subgraph $G_s$, subspace dimension $l$, and random walk step $k$}
    \Ensure {local spectra $\mathbf{V}_{k,l}$}
    \State Compute normalized adjacency matrix $\mathbf{ \bar A}_s$ using (\ref{eqn:normalized_matrix})
    \State Initialize $\mathbf{p}_0$
    \State $\mathbf{V}_{0,l} = \mathtt{orth} (\mathcal{K}_{l+1}(\mathbf{ \bar A}_s,\mathbf{p}_0))$
    \For {$i = 1,...,k$}
    \State $\mathbf{V}_{i,l}\mathbf{R}_{i,l}=\mathbf{\bar A}_s \mathbf{V}_{i-1,l}$
    \Comment {$\mathbf{R}_{i,l}\in \mathbb{R}^{n\times l}$ is obtained by QR factorization so that $\mathbf{V}_{i,l}$ is orthonormal.} 
    \EndFor
  \end{algorithmic}
\end{algorithm}


\subsection{Round Diffusion Vector via Sweeping Cut}

The optimization result $\mathbf{y}$ obtained above is a real-valued vector, where each element $y_i$ hints the propensity for node $i$ to be labeled the same as seed. A commonly adopted method of rounding the diffusion values into labels is to perform a sweep-cut procedure on the nodes ranked by the diffusion value, with an objective of minimizing the graph cut metric such as {\em conductance} \cite{andersen2006local,    mahoney2012local,whang2013overlapping}.

\begin{mydef}
Let $\mathbf{x}\in \{0,1\}^N$ denote the binary indicator vector for the subset $\mathcal{V}'\subseteq \mathcal{V}_s$ and $\mathbf{H}\in \mathbb{R}^{N\times N}$ is any symmetric matrix. The Rayleigh quotient with respect to $\mathbf{H}$ is expressed as the quadratic form of 
\begin{equation}
\rho_{\mathbf{H}}(\mathbf{x}) =\frac{\mathbf{x}^T\mathbf{H}\mathbf{x}}{\mathbf{x}^T\mathbf{x}}.
\end{equation}
\end{mydef}
In particular, conductance of the set $\mathcal{V}'$ measures the fraction of edges leaving $\mathcal{V}'$ among all the edges incident on $\mathcal{V}'$, and can be expressed using a generalized Rayleigh quotient 
\begin{equation}
\Phi(\mathcal{V}') = \rho_{\mathbf{L}_s,\mathbf{D}_s }(\mathbf{x}) =\frac{\mathbf{x}^T\mathbf{L}_s\mathbf{x}}{\mathbf{x}^T\mathbf{D}_s\mathbf{x}} = \frac{\mathbf{x}^T(\mathbf{D}_s-\mathbf{A}_s)\mathbf{x}}{\mathbf{x}^T\mathbf{D}_s\mathbf{x}}.
\end{equation}
We label each node in $\mathcal{V}_s$ by first ranking nodes in decreasing order based on the corresponding value in $\mathbf{y}$. For each prefix set of node $\mathcal{V}' (|\mathcal{V}'| \ge n)$ in the sorted list, we then compute
the conductance of that set and return the set that achieves the minimum, i.e.,  
\begin{equation}
\mathcal{C}' = \argmin_{\mathcal{V}'\subseteq \mathcal{V}_s} \Phi(\mathcal{V}')
\end{equation}
\begin{mylem}
The set $\mathcal{C}'\subseteq \mathcal{V}_s$ found by the local spectral diffusion method is an $[n,m,\frac{\mathbf{x}^T\mathbf{A}_s\mathbf{x}}{\mathbf{x}^T(\mathbf{J-I})\mathbf{x}},\Delta t]$-temporally approximate bipartite core, where $\mathbf{x}\in \{0,1\}^N$ is the binary indicator vector for $\mathcal{C}'$ and $\mathbf{J} \in \mathbb{R}^{N\times N}$ is a matrix of all ones.
\end{mylem}

\textsc{Proof.}
The constraint of $|\mathcal{C}'| \ge n$ is automatically met by the sweeping cut procedure. The fact that $G_s$ only consists of edges with weight greater than $m$ also ensures the constraint specified in $(5)$. Furthermore, $\frac{\mathbf{x}^T\mathbf{A}_s\mathbf{x}}{\mathbf{x}^T(\mathbf{J-I})\mathbf{x}}$ is the quadratic equivalence to the internal density measurement of a cluster, i.e., $2|\mathcal{E}'|/n(n-1)$. \qed


\section{User Engagement Graph} \label{sec:graph}

\subsection{Graph Builder}
We create engagement graph by using interactions between users to model the way users interact with a video or a channel. This allows us to detect {\em orchestrated} actions by sets of users which have a very low likelihood of happening spontaneously or organically.


In practice, the YouTube Comment engagement graph is built with the anonymized aggregate YouTube user activity logs from the past $30$ days window, and is updated on a daily basis using a MapReduce implementation. Here we take the snapshot of graph created on August 3rd, 2015. The Comment engagement graph consists of hundreds of thousands of nodes and tens of millions of edges. The detailed statistics of the engagement graph in use are not discussed here for privacy reasons. Note that the engagement graph we created here constitutes a subgraph of the entire YouTube engagement graph, where we only captured entities that had activities within the scope of a month.


In the engagement graph, nodes represent users and edges represent common videos or channels on which the users engage. Users that have interacted with a common video will share an edge and are consequently joined in the graph. 
Edge weights are by default computed based on the number of common engagement activities between two nodes. For example, in the case of users commenting on a YouTube video, this approach translates into users having and edge weight between them equal to the number of common videos they have commented  upon.

\vspace{0.5em}
{\bf \noindent Adding weight penalty}
\vspace{0.2em}

\noindent The way we built the YouTube Comments engagement graph is essentially the same as above except for the subtle difference that node can be two types of entities -- a user or a Google\texttt{+} Page. It is worthwhile noting here that YouTube comments can be made through the Google\texttt{+} social platform, without having to log into the YouTube sites. Such feature was powered by YouTube's Google\texttt{+} comment integration system introduced in November, 2013. Each PlusPage behaves like a unique user ID and can be used to write comments across platforms including YouTube.


\begin{figure}[htbp]
\begin{center} \includegraphics[width=0.5\textwidth]{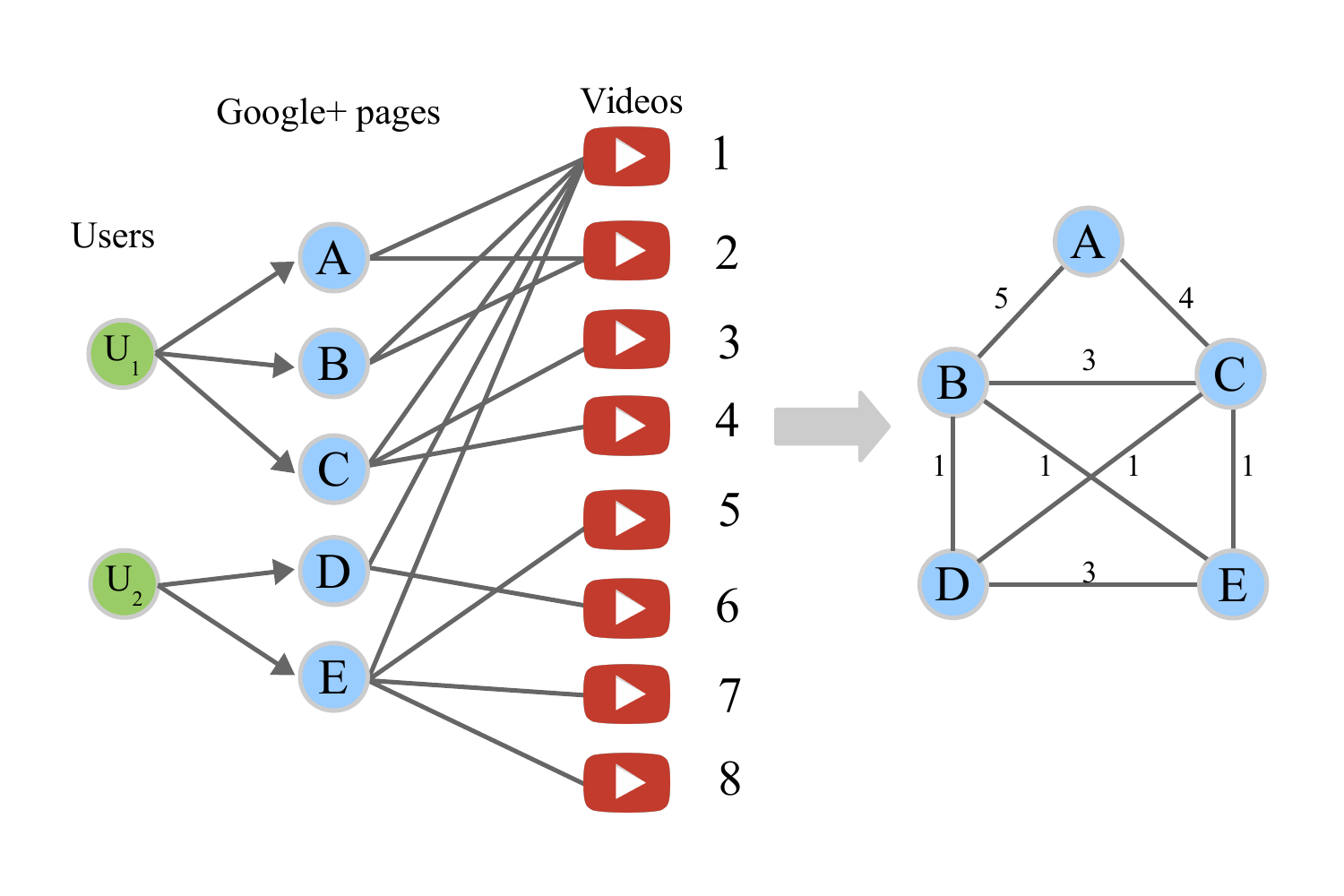} \end{center}
\caption{\bf Example of constructing Google\texttt{+} pages engaged graph. It shows a group of two users using their PlusPages to spam video \texttt{\#}1.} 
\label{fig:graph_builder}
\end{figure}

In order to detect abuse originating from PlusPages, we add an additional step when constructing the graph. This modification tend to penalize those PlusPages created by the same user the following way:
\begin{equation}
\tilde w_{p_i,p_j} = \mathbbm{1}({u(p_i) = u(p_j)}) \cdot |\mathcal{P}(u(p_i))|  + w_{p_i,p_j},
\end{equation}
where $\mathbbm{1}(\cdot)$ is the indicator function; $u(\cdot)$ defines the owner of a PlusPages and $\mathcal{P}(\cdot)$ gives the set of PlusPages a user has created. We use $w_{p_i,p_j}$ to denote the original edge weight between PlusPages $p_i$ and $p_j$, and is calculated by the number of common videos both $p_i$ and $p_j$ commented on. $\tilde w_{p_i,p_j}$ is the updated edge weight, and is equal to $w_{p_i,p_j}$ when $p_i$ and $p_j$ share different owners. In the case where $p_i$ and $p_j$ are created by the same user, we add extra weight regulated by the total number of PlusPages the user has created. The rationale being that owning a larger number of PlusPages indicates a stronger signal of being potentially abusive.

Figure \ref{fig:graph_builder} gives an example of constructing Google\texttt{+} pages engaged graph. It shows that the edge weight between $(A,B)$, $(A,C)$ and $(B,C)$ are all increased by $3$, which is the total number of PlusPages the user $U_1$ has created. The clique structure formed by node $A, B$ and $C$ becomes more noticeable after applying the penalty.
\subsection{Spammer Seeds}

In the context of anomaly detection, when we find suspicious users, we often want to quickly find additional users with similar patterns of behavior that should be disabled as well.  \textsc{Leas} makes use of those users that are identified to be abusive from other YouTube's security mechanisms as seeds. 

In practice, spammer seeds are also updated on a daily basis together with the engagement graph. Since the number of available seeds can be limited, \textsc{Leas} can greatly expand the coverage of daily fake engagement take-down volume.


\begin{figure}[t]
\centering
\begin{subfigure}{\columnwidth}
\includegraphics[width=\columnwidth]{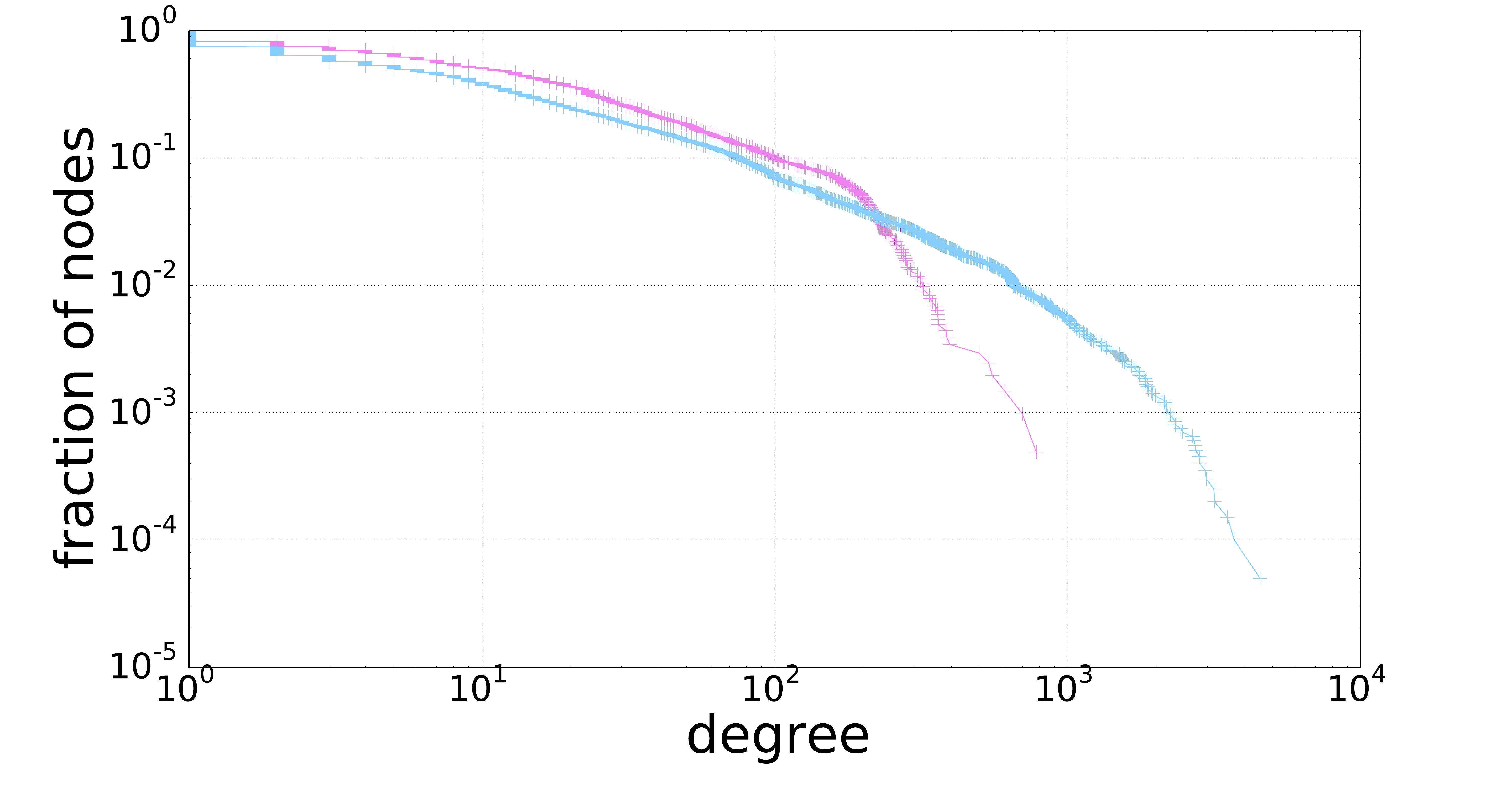}%
\label{subfiga}%
\end{subfigure}\hfill%
\caption{\bf Comparison of node degree distribution between spammers and the general population in YouTube Comment engagement graph. The degree distribution of seeds is depicted in magenta, whereas the distribution of general population is depicted in blue. The number of seeds used for plotting is 2k.  To plot the general population distribution, we first randomly sampled 10k nodes from the engagement graph. We further excluded those known abusive nodes from the sampled population, which left us with 9,957 nodes. Note that the sampled population may contain unknown malicious nodes.}
\label{fig:degree}
\end{figure}

\vspace{0.5em}
{\bf \noindent Degree distribution}
\vspace{0.2em}

\noindent We started probing into the behavior pattern between the spammer nodes and the general population by examining the node degree distribution.  A salient observation from Figure \ref{fig:degree} is that the degree distribution of seeds (depicted in magenta) has a dissimilar tail effect compared to that of the general population (depicted in blue). And the difference can be been across all engagement-level activities, and it is the most evident in the Comments graph. 


This observation surprisingly corresponds with the fact that spam campaigns and companies involved in selling fake engagements may have efforts in relatively modest scope and scale. For example, we looked into several existing online vendor sites that claim to sell YouTube fake engagement. Through investigation we found that 
YouTube comments are usually sold with package size ranging from 15 to several hundred, which matches exactly with the seed degree distribution in Figure \ref{fig:degree}. For example, we find spammer nodes rarely have degree greater than 781 in the Comments graph.


\vspace{2em}
\section{A MapReduce Implementation} \label{sec:mapreduce}

Our local spectral diffusion method enables a straightforward adaption to the MapReduce implementation framework. In this Section, we introduce practical details and also potential caveats in applying the method at scale. The implementation is provably scalable to massive datasets and trivially parallelizable, with the capability of searching for many clusters simultaneously. Furthermore, our pipeline has the same performance guarantee as the serialization since each diffusion procedure is performed locally on the graph. 

{\bf Data Server} The engagement graph is served using SSTableService, a distributed in-memory key-value serving system within Google. Each data server holds a partition containing $1/P$ of the total amount of data, where $P$ denotes the number of shards (partitions) of the data. SSTableService allows serving graph queries in a much faster speed compared to on-disk queries. The SSTableService is shared across mappers when running the job. 

{\bf Data Format} We use {\em Protocol Buffers}\footnote{\url{https://developers.google.com/protocol-buffers/}} for defining the I/O data streams in our implementation. Each protocol buffer message is a small logical record of information, containing a series of name-value pairs. The $\mathtt{graph}$ protocol namely stores the weighted adjacency list keyed by each node; the $\mathtt{seed}$ protocol contains the IDs of the spammer seeds; and the $\mathtt{accomplice}$ protocol defines the output of detected accomplice clusters consisting of suspicious nodes with similar pattern of behavior as the seed. Additionally, we define $\mathtt{config}$ protocol for conveniently encapsulating and passing configuration parameters to each mapper when initializing the jobs. Some tunable parameters in our pipeline include, for example, the dimensionality of local spectral subspace $l$, the number of short random walk steps $k$, 
the minimum cluster size $n$, the maximum size of the sampled subgraph $N$, the degree threshold $d_\text{max}$ for sampling the subgraph, the edge weight threshold $m$.
\vspace{1em}
\floatname{algorithm}{Algorithm} 
\renewcommand{\algorithmicrequire}{\textbf{Globals:}$\mathtt{graph}~G=(\mathcal{A},\mathcal{E},\mathcal{W})$, configuration parameters}
\begin{algorithm} \label{alg:mapreduce}
  \caption{\textsc{MapReduce}~\textsc{Leas}}
  \begin{algorithmic}[1]
    \Require{}
    \State \textsc{InitializeReplica()}
  	\For {$s\in \mathcal{S}$}
  	\If {$\mathtt{deg}(s) \le d_{\text{max}}$}
  	\State Sample subgraph $G_s$
  	\State $\mathbf{V}_{k,l}$ = $\textsc{Local}\textsc{Spectral}(G_s,s)$
  	\Comment {compute local spectral subspace}
  	\State {Solve the optimization objective $\mathbf{y}$ in Section \ref{section:mon}}
  	\State {$\mathcal{C}' = \textsc{Sweep}\textsc{Cut}(\mathbf{y})$}
  	\State \textbf{emit} $\langle s, \mathtt{accomplice}~\mathcal{C}' \rangle$
  	\EndIf
  	\EndFor 
  \end{algorithmic}
\end{algorithm}

\begin{figure*}[t] 
\begin{center} \includegraphics[width=\textwidth]{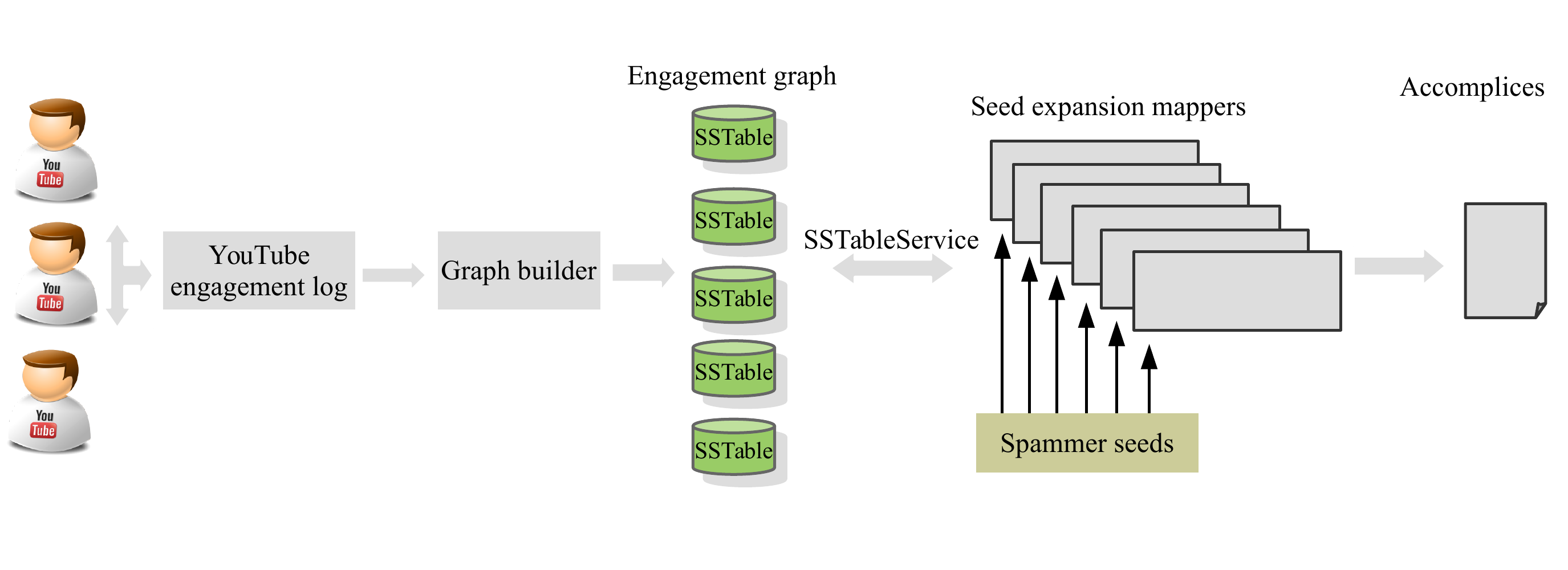} \end{center}
\vspace{-1.5em}
\caption{\bf MapReduce implementation of YouTube fake engagement detection pipeline.} 
\label{fig:pipeline}
\end{figure*}

The core of the MapReduce \textsc{Leas} algorithm can be seen in Algorithm 2. The module of \textsc{InitializeReplica} passes the parameters defined by the configuration protocol to all the mappers. And each mapper job processes one seed at a time independently. The entire pipeline of fake engagement detection is illustrated in Figure \ref{fig:pipeline}, which encompasses the main components of graph builder and seed expander. The graph builder is also implemented using MapReduce framework, where the details are omitted here due to space limit.

\vspace{2.5em}
\section{Experimental Analysis} \label{sec:experiments}

\subsection{Scalability}

\noindent YouTube now has over a billion users and is continuing to grow. Therefore, it is important for the algorithm scales well to large datasets in order to efficiently catch the fake engagement activities on a daily basis. We test and compare the performance with \textsc{CopyCatch}, which is the state-of-the-art algorithm that detects fake Page Likes by analyzing the engagement graph of user-Page interaction. 

\begin{figure*}%
\centering
\begin{subfigure}{\columnwidth}
\includegraphics[width=\columnwidth]{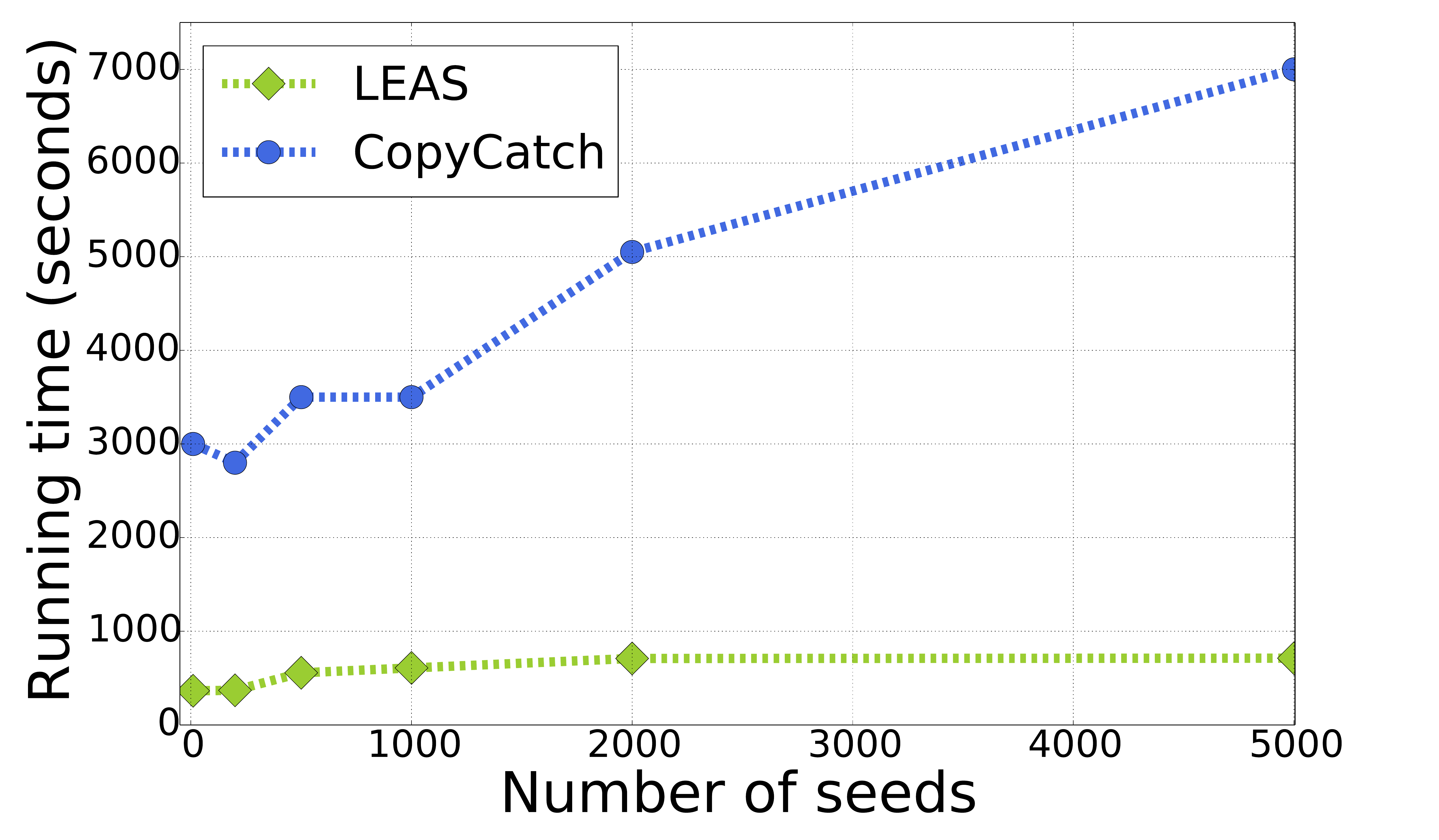}%
\caption{}%
\label{fig:running_time}%
\end{subfigure}\hfill%
\begin{subfigure}{\columnwidth}
\includegraphics[width=\columnwidth]{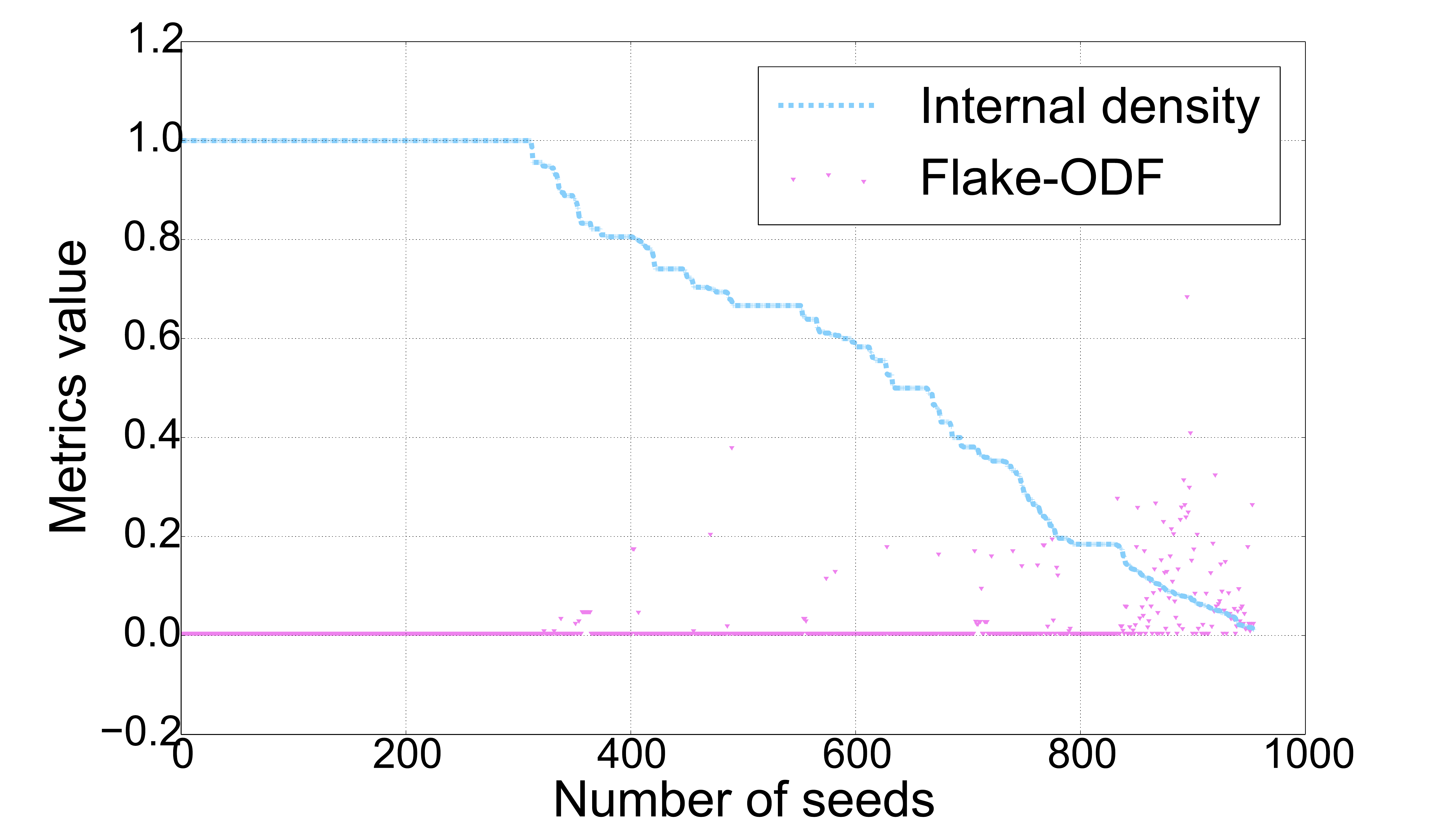}%
\caption{}%
\label{subfiga}%
\end{subfigure}\hfill%

\caption{\bf (a) Comparison of pipeline running time with state-of-the-art as the number of seeds increases. (b) Internal density and Flake-ODF of detected accomplice clusters in YouTube Comments engagement graph. We filtered those seeds with degree greater than 500, i.e., $d_{\text{max}=500}$ and performed the diffusion algorithm on the rest of the seeds. The number clusters in plot is 955. Cluster indices are sorted by the internal density value.}
\label{fig:metric}
\end{figure*}

Firstly, we test the scalability of the algorithm by running our implementation on the YouTube Comments graph over different number of seeds. To make the test results comparable, we choose the same set of seed numbers as that reported in \cite{beutel2013copycatch}. The number of seeds varies from 100 to 5,000. We additionally run the pipeline with only 10 seeds to test the system starting-up time. Depending on the resources availability, it usually takes about $4\sim6$ minutes for the system to allocate and set up the data servers and the MapReduce clusters. Figure \ref{fig:running_time} shows the comparison of running time between \textsc{CopyCatch} and \textsc{Leas}\footnote{We refer to the experimental results originally reported in \cite{beutel2013copycatch} for evaluation.}. It is worthwhile noting that \textsc{Leas} achieves 10 times faster running time with much fewer machines. For example, 3,000 mappers and 500 reducers were used for all the testing data points in \cite{beutel2013copycatch}, whereas at most 1,500 mappers and 2 reducers are required in \textsc{Leas} test run with 5,000 seeds. Even fewer mappers are required for those tests with smaller number of seeds. 
For example, running the pipeline with 1,000 seeds uses 295 mappers, 2,000 seeds uses 597 mappers and 10,000 seeds uses 2,999 mappers. 

As seen in the results, we find that the running time of \textsc{Leas} is almost independent of the number of seeds. This is reassuring that our implementation exploits the parallelism of the problem and can continue to scale as the data scales.

\subsection{Performance Evaluation}
\subsubsection{Graph Metrics}

To evaluate the accomplice clusters found by \textsc{Leas}, we first measure the structural properties using two commonly adopted metrics \cite{yang2015defining}. 
\begin{itemize}
\item {\bf Internal density}  measures the internal edge density of a node set $\mathcal{V}'$. A larger internal density value indicates a more densely connected community-like structure among nodes. 
\begin{equation*}
f(\mathcal{V}') = \frac{2|\mathcal{E}'|}{|\mathcal{V}'|(|\mathcal{V}'|-1)}
\end{equation*}
\item {\bf Flake-ODF} is a cluster metric that takes into account both the internal and external connectivity of a set. It measure the fraction of nodes in $\mathcal{V}'$ that have fewer edges pointing inside than to the outside of the set. Ideally, a smaller Flake-ODF value indicates a better cluster quality. 
\begin{equation*}
f(\mathcal{V}') = \frac{ |\{v:v\in\mathcal{V}', | \{(v,u)\in \mathcal{E}': u\in \mathcal{V}'\}| < \mathtt{deg}(v) / 2\}|} {|\mathcal{V}'|}
\end{equation*}
\end{itemize}

Figure \ref{fig:metric} presents the measurement scores of accomplice clusters detected in three YouTube Comments engagement graph. The most striking observation is the difference concerning the internal density distribution exhibited by the Comments graph. We see that clusters detected from the engagement graph in general are compact with high  internal density, which may signify the orchestration strategy when performing fake engagement --- that the YouTube fake Comments spammers are exposed to have stronger lockstep behavior pattern, where groups of users acting together, commenting on the same videos at around the same time. The clusters corresponding to the tail part of the curve, on the other hand, displays a less orchestrated pattern with more likelihood to be incentivized campaigns. Our probe into the structural properties of the detected clusters also suggests that further evaluation  is imperative. 



\subsubsection{YouTube Comment: Manual Review Results}

\begin{figure}
\begin{center} \includegraphics[width=0.5\textwidth]{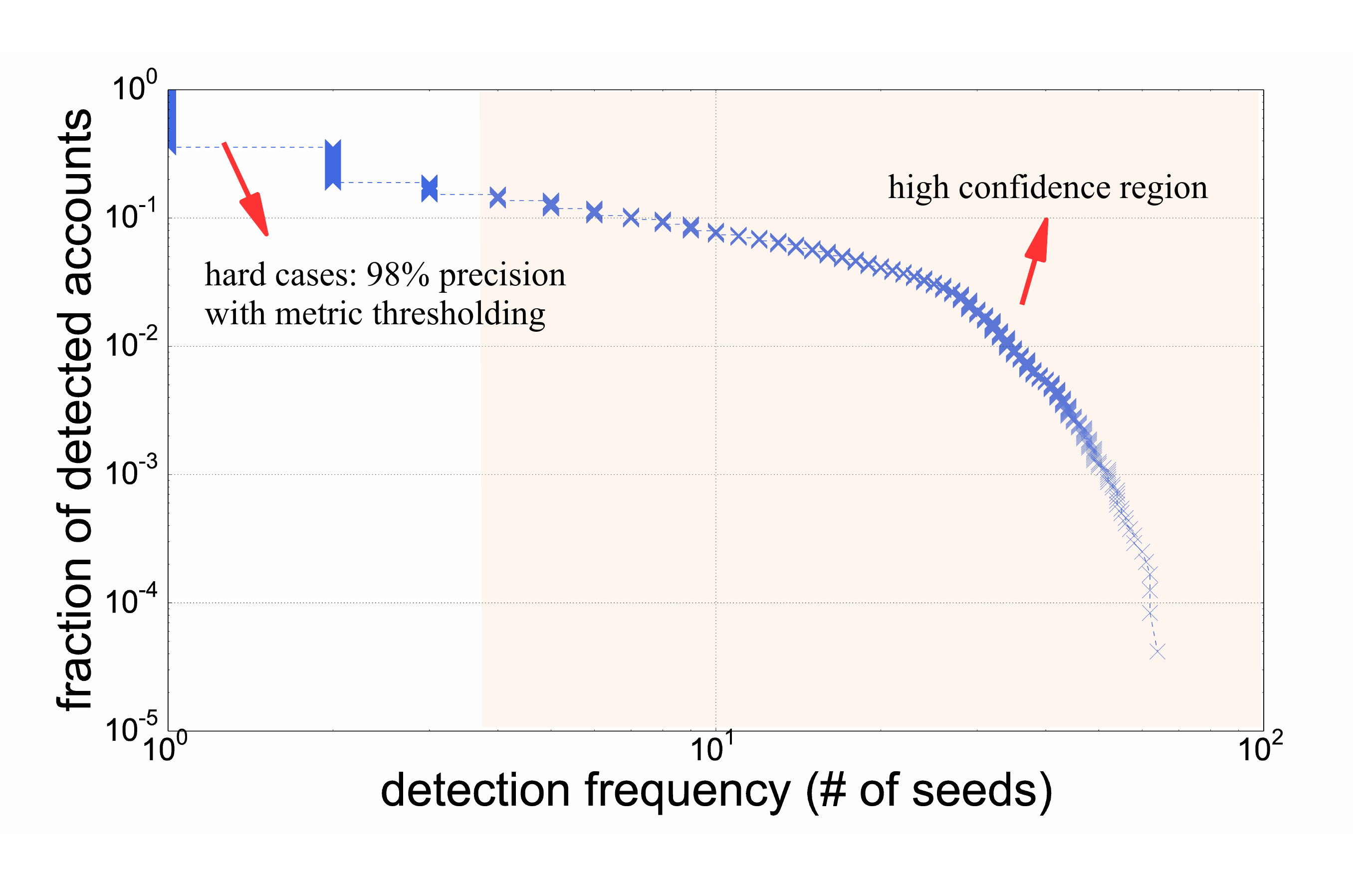} \end{center}
\caption{\bf Detection frequency distribution of among the accounts detected by LEAS.} 
\label{fig:comment_result}
\end{figure}

To verify the effectiveness of the algorithm, we ran the pipeline on the engagement graph built on August 3rd, 2015 within 30 days of time window, and performed intensive manual review on the detected accounts. In total, the pipeline detected roughly 24,000 unique accounts with 955 spammer seeds. Among the newly detected accounts, we find that 8,500 of them are found by more than one seed; while the other 15,500 accounts are detected by only one seed. Figure \ref{fig:comment_result} depicts the distribution of the frequency for each account being detected by certain seed(s). The fact that an account detected by several seeds is a stronger indication of being potentially abusive. We therefore divide the results into two types and perform analysis accordingly:
\begin{itemize}
\item {\bf Tier I:} accounts that are repeatedly detected by more than one seed ($35\%$). 
\item {\bf Tier II:} accounts that are uniquely detected by only one seed ($65\%$). 
\end{itemize}    

\begin{figure*}[htbp]%
\centering
\begin{subfigure}{1.05\columnwidth}
\includegraphics[width=\columnwidth]{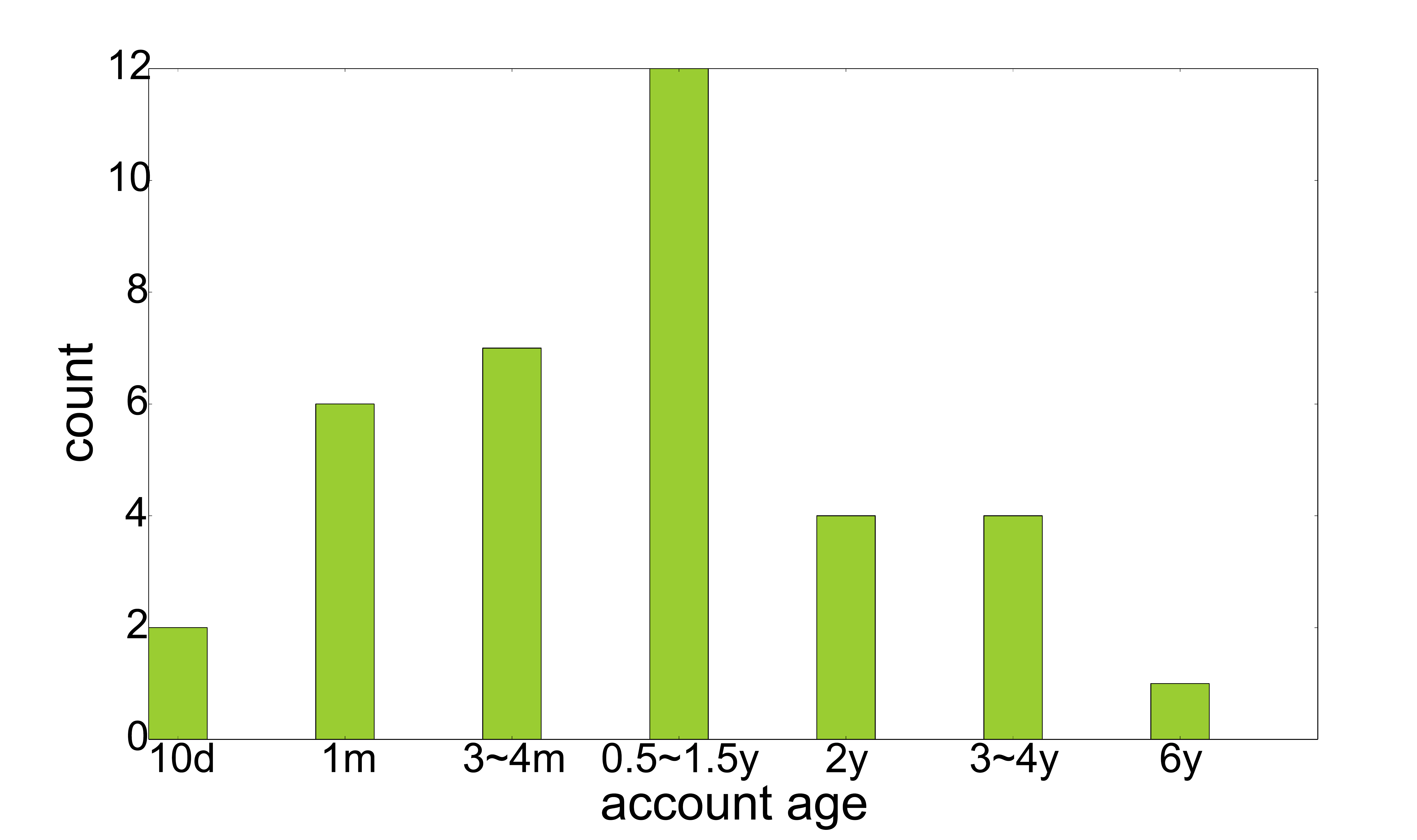}%
\caption{}%
\label{fig:age}
\end{subfigure}\hfill%
\begin{subfigure}{0.95\columnwidth}
\includegraphics[width=\columnwidth]{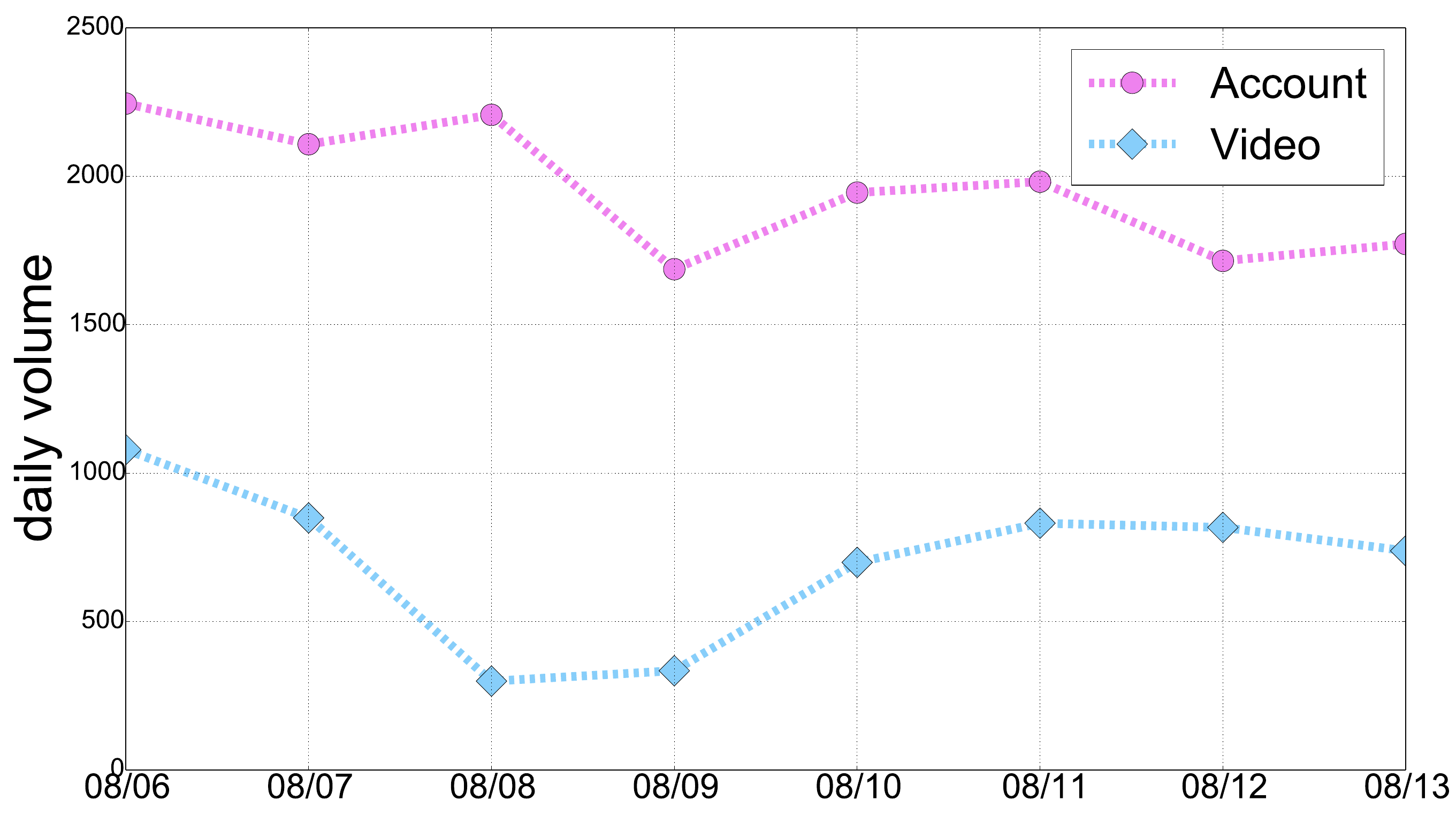}%
\caption{}
\label{fig:aggregate}
\end{subfigure}\hfill%
\caption{
\bf (a) Age distribution of 36 manually reviewed Tier I suspicious accounts. (b) Google live runs on YouTube engagement graphs with portion of the seeds, dating from August 6th to August 13th, 2015. The magenta curve depicts the daily volume of unique accounts detected by LEAS pipeline, and the blue curve indicates the daily number of videos these accounts have acted upon.}
\end{figure*}

To investigate the Tier I accounts, we randomly selected 36 accounts without applying any metric thresholding. We manually examined each account's information and YouTube post history. We also take into consideration the Google internal security measures associated with each account, but will not discuss in detail here for security reasons. The manual review shows that 100\% of the Tier I accounts were verified to be fake. Among the Tier I accounts, the most frequently detected account was found by 64 seeds. We find that this particular account was created less than 10 days ago yet had posted more than 253 posts with many quota exceeded. We manually clicked through the comments posted by these accounts, and found that most comments are short text pieces such as ``{\em good videos}'', ``{\em very cool }'', ``{\em nice}", ``{\em oh}", ``{\em lol}" or emoji of smile faces. We also find the common pattern for accounts to  post exactly the same or similar short, fake comments to different videos. Besides, we also discovered a few accounts posting comments under popular songs, the contents of which are irrelevant to the video content itself but rather asking for view and subscribe (e.g., ``{\em please subscribe}" or ``{\em subscribe now}"). Additionally, several other spammy accounts posting comments including malicious URLs and advertisement were detected.

Besides the contextual information, we also looked into the lifespan of each suspicious account. Although one might expect most spammer accounts to have relatively young age, it was actually quite surprising to see the age heterogeneity of those accounts, as shown in Figure \ref{fig:age}.
Among the 36 accounts, the most frequent age falls into the range between 0.5 and 1.5 years; whereas the oldest spammer account have already been existent for more than 6 years.

The Tier II accounts are the harder cases. In order to guarantee the FP guards in production, we randomly selected 100 Tier II accounts that belong to an accomplice cluster with internal density greater than 0.7. The  manual investigation shows that 98\% detected Tier II accounts to be fake\footnote{In practice, we treat activities made by both Tier I and Tier II accounts as fake engagement. }. The comments posted by these accounts share similar pattern as those made by Tier I accounts. Quite interestingly, we indeed found a detected cluster of 15 accounts posting the same comments of either ``{\em i love pets}", ``{\em yeah}" or URLs under certain videos. This further verified that the suspicious groups detected by the algorithm are of high accuracy. As for the other two accounts we are uncertain about, one has huge amount of Google\texttt{+} shares of good deals although it posted nothing on YouTube comments; another 4-month old account posts a mixture of both organic and fake-like comments, which might be incentivized.

\subsection{Deployment at Google}
\textsc{Leas} now runs regularly at Google, expanding the coverage of fake engagement activities on YouTube. Parameters have been chosen to significantly distinguish organic user behavior from fake social engagement. There are two levels of take-down actions in practice --- engagement level and account level. Engagement level take-down is a soft penalty which removes all the fake engagement activities happened during the day associated with the detected accounts; account level take-down is a more severe outcome, which is applied when we have very high confidence in certain bad actors committing fake engagement from time to time. Figure \ref{fig:aggregate} shows the daily aggregate volume of detected accounts when running our pipeline on YouTube Comments graphs with {\em portion} of the spammer seeds, dating from August 6th to August 13th, 2015\footnote{The decreased amount of detected account on August 8th and 9th was due to the reduced number of available seeds. In practice, the seeds data is provided by YouTube abuse team and the quantity of which may vary from day to day.}. We do not display the entire daily take-down volume here for security reasons. Note that engagement level take-down was the main penalty applied during our test runs, henceforth the detected accounts didn't exhibit a fluctuation from day to day --- otherwise we would expect to see a decreasing volume of detected accounts  when applying the account level take-down policy. Overall, this method, in combination with other existing abuse infrastructure at Google, is effective in decreasing the volume of fake social engagement on YouTube.


\section{Conclusion and Extensions} \label{sec:conclusion}
In this paper, we show how fake social engagement activities on YouTube can be tracked over time by analyzing the temporal engagement graph, which models the interactions between users and YouTube video objects. With the domain knowledge of spammer seeds, we formulate and tackle the problem of detecting fake social engagement in a semi-supervised manner --- with the objective of searching for individuals that have similar pattern of behavior as the known seeds --- based on a graph diffusion process via local spectral subspace. We show our method, \textsc{Leas}, is scalable to massive datasets, with a straightforward adaption to the MapReduce implementation. We demonstrate the effectiveness of our deployment at Google by achieving a manual review accuracy of 98\% on YouTube Comments graph in practice. Our examination on the anonymized YouTube log data also revealed multitudes of different patterns of behavior between abusive accounts and the general population, measured by the average co-engagement intensity, monthly aggregate activity, for instance. 

By clustering YouTube users based on their engagement behavior pattern, \textsc{Leas} has shown to greatly expand the coverage of daily fake engagement take-down volume on YouTube. Our approach can be extended to many other settings including Twitter followers, Amazon product reviews and Facebook Likes etc. We envision two future directions towards which our work can evolve. First, while this paper describes the approach in a generic setting, our method can be extended by incorporating other meta signals such as IP address the engagement activities were made from. Second, we believe that better detection model can be derived by taking into account the incentivized engagement behavior, where users are offered incentives (e.g. bonus point rewards) to act on a target such as writing product reviews.

\section{Acknowledgements}
The authors would like to thank Google YouTube abuse team for providing the valuable YouTube Comments data and spam seeds. Yixuan Li has been supported by US Army Research Office W911NF-14-1-0477. 




%
\vspace{0.5em}
\bibliographystyle{abbrv}
{\bibliography{sigproc}}
%
%

\end{document}